\documentclass{iopart}

\usepackage{cite}                       
\usepackage[dvips]{epsfig}

\begin{document}
\title{Hole crystallization in semiconductors}
\author{
M.~Bonitz$^{1}$,
V.S.~Filinov$^{1,2}$,
V.E.~Fortov$^{2}$,
P.R.~Levashov$^{2}$,
H.~Fehske$^{3}$}

\address{$^{1}$ Christian-Albrechts-Universit{\"a}t  Kiel,
Institut f{\"u}r Theoretische Physik und Astrophysik, 24098 Kiel, Germany}
\address{$^{2}$ Institute for High Energy Density, Russian Academy of Sciences,
Izhorskay 13/19, Moscow 127412, Russia}
\address{$^{3}$ Universit{\"a}t Greifswald, Institut f{\"u}r Physik, l7487 Greifswald, Germany}

\begin{abstract}
When electrons in a solid are excited to a higher energy band they leave behind a 
vacancy (hole) in the original band which behaves like a positively charged particle. Here we predict that holes can spontaneously order into a regular lattice in semiconductors with sufficiently flat valence bands. The critical hole to electron effective mass ratio required for this phase transition is found to be of the order of 80. 
\end{abstract}

\noindent {\bf Motivation.}
More than seven decades ago Wigner predicted the existence of a crystalline state  of the electron gas in metals at low densities -- the electron Wigner crystal \cite{wigner}. Since then, there has been an active search for this strong correlation phenomenon in many fields. Finally, crystallization of electrons was observed on the surface of helium droplets \cite{grimes}, and it is predicted to occur in semiconductor quantum dots \cite{afilinov-etal.01prl}. There are also predictions of electron crystallization in semiconductor heterostructures in the presence of a strong magnetic field (which acts in favor of electron localization) but there is so far no conclusive confirmation. The necessary condition for the existence of a crystal in these {\em one-component plasmas} (OCP) is that the mean Coulomb interaction energy,
$e^2/{\bar r}$ (${\bar r}$ denotes the mean inter-particle distance), exceeds the mean kinetic energy (thermal energy $\frac{3}{2}k_BT$
or Fermi energy $E_F$ in classical or quantum plasmas, respectively) by a large factor
$\Gamma^{cr}$ which, in a classical OCP is given by $175$ \cite{grimes,dewitt}. In a quantum OCP at zero temperature the coupling strength is measured by the Brueckner parameter,
$r_s\equiv {\bar r}/a_B$ ($a_B$ denotes the effective Bohr radius), the
critical value of which is $r_s^{cr}\approx 100$ \cite{ceperley80}.

On the other hand, Coulomb crystallization has been observed in {\em neutral two-component plasmas} (TCP), e.g. in colloidal and dusty plasmas \cite{thomas94,hayashi94,arp04,ludwig-etal.05pre}. Besides these {\em classical TCP crystals} it is thought that in the interior of white dwarf stars and in the crust of neutron stars there exist crystals of bare carbon, oxygen and iron nuclei which are embedded into an extremely dense degenerate Fermi gas of electrons, see e.g. \cite{segretain}. No such {\em quantum TCP crystals} have been observed in the laboratory, despite early suggestions \cite{rice68}. It is, therefore, of high 
interest to analyze the necessary conditions for the existence of Coulomb crystals in 
a two-component plasma to understand in which other TCP systems crystallization is 
possible, which is the aim of the present paper. 

\begin{figure} [htp]
\centering
\includegraphics[scale=1,angle=-0]{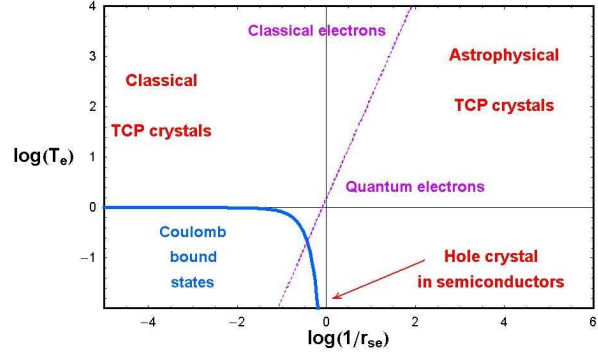}
\vspace{-0.4cm}
\caption{(Color online) Location of the classical and quantum TCP crystals and of the hole 
crystal in the density-temperature plane (qualitative picture).}
\label{phase0}
\end{figure}

\noindent {\bf Criterion for the occurence of a hole crystal.}
A qualitative phase diagram which 
shows the location of the mentioned TCP crystals is shown in Fig.~\ref{phase0}.
Note that these Coulomb crystals are very different from the common crystals observed 
in classical ionic systems like salts (e.g. NaCl) or metals which cannot be described 
in terms of a simple coupling parameter. The properties of the latter systems 
depend on the microscopic structure of the ionic constituents (electronegativity,
in the case of salts, and band structure, in the case of metals etc.). Here we will 
concentrate on plasma-like systems which involve pointlike ions (not containing deeply  bound electrons). 
\vspace{-0.4cm}

We consider a locally neutral macroscopic system of electrons (e) and holes (h). 
The equilibrium state is characterized by the dimensionless electron temperature
$T_e = 3 k_BT/2 E_B$ and the mean inter-electron distance $r_{se}={\bar r}_e/a_B$,
where $E_B=\frac{e^2}{4\pi\epsilon_0\epsilon_r}\frac{1}{2a_B}$ denotes the exciton
 binding energy, $a_B =  \frac{\hbar^2}{m_r}\frac{4\pi\epsilon_0\epsilon_r}{e^2}$ 
is the exciton Bohr radius, $\epsilon_r$ and $m_r$ are the background dielectric constand and the reduced mass $m_r^{-1}=m_h^{-1}(1+M)$]. The
dimensionless density is given by $na_B^3=3/(4\pi r^3_{se})$. In addition to these parameters which also characterize an OCP, the state of the electron-hole 
plasma is characterized by the mass ratio $M=m_h/m_e$. 

The condition for a hole crystal in a TCP follows from the OCP crystal condition,
$r_{sh} \ge r_s^{cr}$, after rescaling ${\bar r}$ and $a_B$ by
taking into account the charge and mass ratio yielding $(M+1)r_{se}\ge r_s^{cr}$.
\vspace{-0.5cm}
\begin{figure} [htp]
\centering
\includegraphics[scale=0.29,angle=-90]{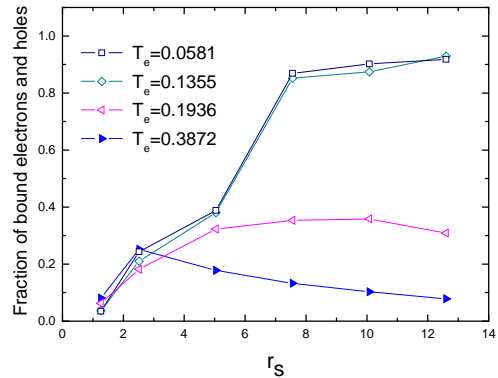}
\vspace{-0.5cm}
\caption{(Color online) PIMC simulation results for the electron-hole bound state fraction (including excitons and biexcitons) in a $3d$ semiconductor ($M=40$) vs. inverse density for several temperatures given in the figure.}
\label{alpha}
\end{figure}
This Coulomb crystal of holes will survive in the presence of electrons only if  
holes do not form bound states, as this would drastically reduce the correlation energy 
of the holes, thus eventually reducing the coupling strength below the critical level necessary for crystallization. At zero temperature, bound states break up due to pressure ionization at densities above the Mott density, 
i.e. $r_{se} \le r_s^{{\rm Mott}} \approx 1.2$. This has been confirmed by first principle path integral Monte Carlo (PIMC) simulations \cite{fehske-etal.05jp}. As 
the numerical data in Fig.~\ref{alpha} confirm, at $r_{se}\le 1.2$ less than $10\%$ of the holes are bound in excitons and biexcitons.
With increasing temperature, ionization becomes possible at lower density which we indicate by a monotonically decreasing function $1/r_s^{{\rm Mott}}(T_e)$ which
vanishes when $T_e \rightarrow 1$ because there thermal ionization prevails.

\noindent {\bf Critical mass ratio for the hole crystal.}
Combining the above expression for $r_{se}$ with the existence of pressure ionization
yields the {\em criterion for the existence of a TCP crystal} in the presence of
a neutralizing background of {\em quantum electrons} as
\begin{eqnarray}
M \ge M^{cr}(T_e) = \frac{r_s^{cr}}{r_s^{{\rm Mott}}(T_e)} - 1,
\label{mcr}
\end{eqnarray}
which exists in a finite electron density range $[n^{(1)},n^{(2)}]$ given by
\begin{eqnarray}
n^{(1)}(T_e) &=&\frac{3}{4\pi}\left[\frac{1}{r_{se}^{\rm Mott}(T_e)}\right]^3,
\quad
n^{(2)}(T_e) = n^{(1)}(T_e)K^3, 
\label{nmax}
\end{eqnarray}
where $K= (M+1)/(M^{cr}+1)$. The crystal exists below a maximum temperature $T^*$, which is estimated by the crossing point of the classical and quantum asymptotics of an OCP crystal \cite{afilinov-etal.01prl}
$T^{*} = 6 (M+1)/(\Gamma^{cr}r_s^{cr})$.
According to Eq.~(\ref{mcr}), the critical hole to electron mass ratio is given by $83$ 
at zero temperature. This value decreases with increasing temperature (due to the lower Mott density). 

Of course, the critical mass ratio and the density and temperature limits carry some uncertainty arising from the uncertainty of the Mott density and the  critical value of the Brueckner parameter. In fact the transition from an exciton gas to a hole crystal may involve many intermediate states with liquid-like behavior, e-h droplets (phase separation), e.g. \cite{filinov-etal.03jpa}, an analysis of which is beyond the present work. We estimate that these effects give rise to an uncertainty of the minimum density (Mott density), $n^{(1)}$, of the order of $30\%$. Further, the error of $r_s^{cr}$ is about $20\%$ \cite{ceperley80}, thus 
the critical parameters carry an uncertainty of about $50\%$. For particular systems, more accurate predictions are possible if the Mott parameter $r_s^{{\rm Mott}}$ is known, e.g. from computer simulations.

{\bf Simulation results.} Note that the complex processes of interest pose an extreme challenge to the simulations: They must self-consistently include the full Coulomb interactions, exciton and biexciton formation in the presence of a surrounding plasma, pressure ionization and the quantum and spin properties of electrons and holes. 
We therefore have performed extensive direct fermionic path integral 
Monte Carlo (PIMC) simulations of a $3d$ e-h plasma which are based on our previous results for dense hydrogen-helium plasmas \cite{filinov-etal.00jetpl}, e-h plasmas \cite{filinov-etal.03jpa} and electron Wigner crystallization \cite{afilinov-etal.01prl}. While the so-called sign problem prohibits PIMC simulations of the ground state of a fermion system, here we 
restrict ourselves to temperatures at the upper boundary of the hole crystal phase, 
i.e. $T_e=0.06\dots 0.2$.
Studying mass ratios in the range of $M=1\dots 2000$ and densities corresponding to $r_{se}=0.6 \dots 13$ the simulations cover a large variety of $3d$ Coulomb systems -- from positronium, over typical semiconductors to hydrogen. 
 
\begin{figure}[htp]
\begin{center}
\includegraphics[scale=0.3,angle=-90]{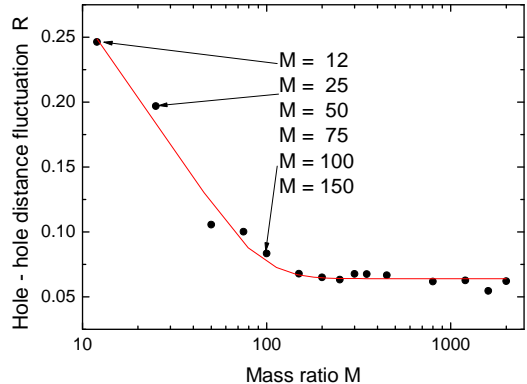}
\end{center}
\vspace{-0.7cm}
\caption{Mean-square relative hole-hole distance fluctuations (normalized to the mean 
interparticle distance) as a function of the mass ratio $M$ for $T_e=0.096$ and $r_s=0.63$. 
Symbols are simulation results, the line is the best fit.}
\label{lindemann}
\end{figure}

Our main results concern the relative distance fluctuations of holes shown in Fig.~\ref{lindemann}. Here we have fixed density and temperature in such a way 
that bound 
state formation is not possible and vary the mass ratio from hydrogen to e-h plasmas. At $M=2000$ the distance fluctuations are small and they remain almost unchanged 
when $M$ is reduced. Around $M=100$ a drastic increase is observed which is a clear 
indication of spatial delocalization of the holes. In fact, analyzing the microscopic 
configuration in the simulation box and the pair distribution functions 
\cite{bonitz-etal.05prl} clearly confirms this interpretation. At large $M$ 
the holes form a crystal which is embedded into a high-density delocalized 
electron gas. This crystal vanishes (melts) between $M=100$ and $M=50$ which is 
in very good agreement with the estimate of Eq.~(\ref{mcr}). 
\vspace{-0.7cm}

The predicted critical mass ratio is larger than in most conventional semiconductors.
However, similar mass ratios have already been reported in intermediate valence semiconductors, such as Tm[Se,Te] \cite{wachter04}. For example, for $M=100$ (using $\epsilon_r=20$) the parameters are $n^{(1)}_e(0)=1.2\cdot 10^{20}{\rm cm}^{-3}$,
$n^{(2)}_e(0)=2.1 \cdot 10^{20}{\rm cm}^{-3}$  and $T^{*} \approx 9K$. Hole 
crystallization could be verified experimentally by means of neutron scattering. 
\\
{\bf Discussion.} Let us briefly mention earlier discussions of the possibility of 
hole crystallization. This effect was first mentioned by Halperin and Rice, e.g. 
\cite{rice68} who mention that the original suggestion is due to C. Herring. First 
rough estimates of the critical hole to electron mass ratio were given by Abrikosov 
who found $M=100$, e.g. \cite{abrikosov78} and $M=50$ in his text book \cite{abrikosov-book}. This estimate is, apparantly, based on assuming a hole lattice 
constant of one exciton Bohr radius. In our first-principle simulations no such assumptions are made, but the results are surprisingly close (we find a maximum lattice constant of about $0.9a_B$ \cite{bonitz-etal.05prl}. Abrikosov also stresses the favorable conditions a hole crystal would have for superconductivity. This is certainly one of the most interesting 
future questions, although our predictions for the critical temperature for the 
hole crystal seem to limit the prospects for high-temperature superconductivity in these 
materials.
\\
{\bf Acknowledgments.} We acknowledge stimulating discussions with H.~DeWitt, M.E.~Fisher 
and P.~Wachter and thank Yu. Lozovik and A.A. Abrikosov for bringing to our attention 
references \cite{abrikosov78,abrikosov-book}. This work is supported by the Deutsche Forschungsgemeinschaft via TR 24 and SFB 652, the RAS program No. 17 and grant No. PZ-013-02 of the US CRDF.
\\

\end{document}